\documentclass[11pt,a4paper,english,showpacs,showkeys,superscriptaddress]{revtex4}

\usepackage{amssymb}
\usepackage{amsmath}
\usepackage{graphicx}
\usepackage{babel}
\usepackage{hyperref}

\usepackage{color}

\begin{document}

\title{Renormalization Group Improvement of the Superpotential for the ${\cal N}=2$ Chern-Simons-matter model}

\author{A.~C.~Lehum}
\email{lehum@ufpa.br}
\affiliation{Faculdade de F\'isica, Universidade Federal do Par\'a, 66075-110, Bel\'em, Par\'a, Brazil.}

\author{J. R. Nascimento}
\affiliation{Departamento de F\'{\i}sica, Universidade Federal da Para\'{\i}ba\\
 Caixa Postal 5008, 58051-970, Jo\~ao Pessoa, Para\'{\i}ba, Brazil}
\email{jroberto,petrov@fisica.ufpb.br}

\author{A. Yu. Petrov}
\affiliation{Departamento de F\'{\i}sica, Universidade Federal da Para\'{\i}ba\\
 Caixa Postal 5008, 58051-970, Jo\~ao Pessoa, Para\'{\i}ba, Brazil}
\email{jroberto,petrov@fisica.ufpb.br}

\author{Huan Souza}
\email{huan.souza@icen.ufpa.br}
\affiliation{Faculdade de F\'isica, Universidade Federal do Par\'a, 66075-110, Bel\'em, Par\'a, Brazil.}


\date{\today}

\begin{abstract}

Within the superfield formalism, we study the renormalization group improvement of the effective superpotential for the ${\cal N}=2$ Chern-Simons-matter theory, explicitly obtain the improved effective potential and discuss the minima of the effective potential and a problem of mass generation in the theory.
\end{abstract}

\maketitle

\section{Introduction}

The effective potential is known to be one of the central objects of quantum field theory allowing to obtain information about low-energy effective dynamics of a theory, spontaneous symmetry breaking, mass generation and other related issues \cite{Coleman:1973jx}. By definition, the effective potential is an effective Lagrangian evaluated at constant background fields, which implied in known difficulties within its generalization to superfield theories where the integral of a constant trivially vanishes. For the four-dimensional superfield theories, the concept of the superfield effective potential has been formulated in \cite{WZ}, where it was argued that for a consistent definition of the effective potential, the background superfield should be constant in a space-time possessing a nontrivial dependence on Grassmannian coordinates. Further, this approach has been successfully applied to study of the effective potential in three-dimensional superfield theories \cite{Ferrari:2009zx}. As a continuation, the one-loop superfield effective potential has been obtained in many three-dimensional theories (see f.e. \cite{ourCS} and references therein). While the interest to the one-loop result is natural since the one-loop contribution is finite in all three-dimensional theories except of those ones with exotic dynamics, it is clear that the study of any theory cannot be restricted by the one-loop order, therefore it is necessary to have some prescriptions for the higher-loop contributions. However, already explicit two-loop calculations in general are rather complicated, moreover, the models representing themselves as couplings of Chern-Simons theory to scalar superfields display divergences in any loop order higher than the first one, cf. f.e. \cite{ourCS}. A powerful tool allowing to obtain the effective potential up to constant multiplier in theories involving divergences is based on use of renormalization group equations (RGE).


The use of RGE in order to improve the calculation of the effective potential has been intensively used in non-supersymmetric theories ~\cite{elias:2003zm,Chishtie:2005hr,PhysRevD.72.037902,Meissner:2006zh,Meissner:2008uw,AGQuinto,Chishtie:2010ni,Dias:2010it,Steele:2012av,Chun:2013soa}, and recently the method has been extended to superspace formalism~\cite{Quinto:2014zaa}. In this paper, we apply this method to ${\cal N}=2$ Chern-Simons-matter theory attracting a strong interest since it represents itself as a simplest $3D$ extended supersymmetric theory interest to which strongly increased in recent years due to studies of three-dimensional AdS/CFT correspondence, see f.e. \cite{BLG}, \cite{BMS} and references therein. As a result, we obtain a generic structure of the one-loop effective potential in our theory.

The paper is structured like follows. In the section \ref{effective_V} we discuss some general features of the effective superpotential in the three-dimensional superspace. In the section \ref{RGE-superspace}, we describe how to use the Renormalization Group Equation to compute the effective superpotential of a classical superconformal theory. In the section \ref{sec_wz} we apply the method for a simple example, the Wess-Zumino model. In the section \ref{n2csm} we describe the classical action of the ${\cal N}=2$ Chern-Simons-matter theory written in the ${\cal N}=1$ superspace. In the section \ref{rge_veff}, we calculate the complete leading log effective superpotential of the ${\cal N}=2$ Chern-Simons-matter theory in terms of a ${\cal N}=1$ background superfield. We also discuss the vacuum structure of the model, where supersymmetry is preserved, showing that no mass is induced by radiative corrections. In the section \ref{summary}, we present a Summary where our results are discussed.

\section{The effective superpotential in the $D=2+1$ superspace}\label{effective_V}

As it is well known, the quantum dynamics of any theory is described in terms of the effective action. While the complete description of the general structure of the supersymmetric effective action in the four-dimensional case is well discussed in \cite{BK0}, this  description requires certain adaptation for the three-dimensional case.

First, the classical action for superfields defined in a three dimensional spacetime can be cast as
\begin{eqnarray}
S=\int d^3x d^2\theta~\mathcal{L}(\Phi, D^\alpha\Phi,D^2\Phi,\ldots)~,
\end{eqnarray}
where $\mathcal{L}$ is the Lagrangian density, $\Phi=\Phi(x,\theta)=\varphi+\theta^{\alpha}\psi_\alpha-\theta^2 F$ is a scalar superfield, $D^2=\frac{1}{2}D^\alpha D_\alpha$ and $D^\alpha$ is the supercovariant derivative. Dots here and below are for terms depending on space-time derivatives of superfields. Throughout this paper, we use the notations and conventions adopted in \cite{SGRS}.

In the three-dimensional superspace, the effective action has a structure that can be presented in a form of the derivative expansion
\begin{eqnarray}
\Gamma[\Phi]=\int d^3x d^2\theta~ \mathcal{V}(\Phi,D_{\alpha}\Phi,D^2\Phi)+\ldots,
\end{eqnarray}
\noindent where the $\mathcal{V}(\Phi)$ is the effective superpotential which depends on the superfield $\Phi$ and its spinor supercovariant derivatives but not on its space-time derivatives.
In general, the $\mathcal{V}(D_{\alpha}\Phi,D^2\Phi)$ can be presented in the form of the derivative expansion:
\begin{eqnarray}
\mathcal{V}(D_{\alpha}\Phi,D^2\Phi;\Phi)=\mathcal{V}_0(\Phi)+\mathcal{V}_2(\Phi)D^2\Phi+\ldots,
\end{eqnarray}
where the $\mathcal{V}_0(\Phi)$ is the zero-order contribution to the effective superpotential, and $\mathcal{V}_{2}(\Phi)D^2\Phi$ is the second-order contribution to it, with the $\mathcal{V}_{2}(\Phi)$ a function of the superfield $\Phi$ but not of its derivatives. The dots correspond to terms with four and more supercovariant derivatives acting over $\Phi$'s. Similarly, for a complex background superfield case, these potentials can depend both on $\Phi$ and $\bar{\Phi}$.

Now, it is necessary to note that the explicit form of the effective potential, at least its contributions of orders zero and two in derivatives,  can be predicted from symmetry and dimension reasons. We assume that the couplings in the theory are dimensionless, just this situation occurs in extended supersymmetric CS models. While in the one-loop order the zero-order contribution to the effective superpotential, by dimensional reasons, must have form $c(\Phi\bar{\Phi})^2$, with $c$ is a constant dependent on couplings in the theory, in higher-loop orders, after subtracting divergences, results like $c(\Phi\bar{\Phi})^2\log^n\frac{(\Phi\bar{\Phi})^2}{\mu^2}$, with any $n\leq L-1$, can arise for any $L$-loop contribution. In principle, however, the renormalization group improvement methodology allows for contributions like $c(\Phi\bar{\Phi})^2\left[\frac{(\Phi\bar{\Phi})^2}{\mu^2}\right]^{\alpha}$, with $\alpha$ is a some number. Similarly, for the second-order contribution to the effective potential we can have $c \bar{\Phi} D^2\Phi \left[\frac{(\Phi\bar{\Phi})^2}{\mu^2}\right]^{\alpha}+h.c.$. Namely the zero and second order contributions will arise further in this paper.

\section{The RGE improvement in the $D=2+1$ Superspace}\label{RGE-superspace}

Let us consider a superconformal field theory in a three dimensional spacetime
\begin{eqnarray}
S=\int d^3x~d^2\theta\left[\frac{1}{2}\Phi D^2\Phi+\frac{\lambda}{4} \Phi^4\right]~,
\end{eqnarray}
where $\Phi=\Phi(x,\theta)=(\varphi+\theta^{\alpha}\psi_\alpha-\theta^2 F)$ is a scalar superfield, $D^2=\frac{1}{2}D^\alpha D_\alpha$ and $D^\alpha$ is the supercovariant derivative.

In order to compute the effective superpotential, we dislocate the superfield $\Phi$ by a constant classical superfield $\phi=\langle\varphi\rangle-\theta^2 \langle F\rangle$ as
\begin{eqnarray}
\Phi(x,\theta)\rightarrow\Phi(x,\theta)+\phi(\theta).
\end{eqnarray}
Then, the classical superpotential $V_{eff}$ can be written as
\begin{eqnarray}
V_{eff}=-\frac{1}{2}\phi D^2\phi-\frac{\lambda}{4}\phi^4,
\end{eqnarray}
where the integration of $V_{eff}$ over the Grassmannian variables result in the effective potential 
\begin{eqnarray}\label{veff_comp1}
U_{eff}=\int d^2\theta V_{eff} &=& -\frac{1}{2} F_{cl}^2-\lambda F_{cl}\varphi_{cl}^3=\frac{\lambda^2}{2}\varphi^6.
\end{eqnarray}
In the last step, we have eliminated out the auxiliary field $F_{cl}$ using its equation of motion to obtain the physical content of the classical effective potential.

In what follows, we use the improved superpotental methodology. Its key idea is as follows -- we start with the  beta functions which in three-dimensional theories began to be contributed at two loops, and solve the corresponding renormalization group equations taking into account that they must be satisfied by the complete effective potential composed by all-loop contributions. The effective potential is naturally expected to be a function of $\log \frac{\phi^2}{\mu}$ and couplings. So, as a result, through analysis of renormalization group equations characterized by these arguments, with the initial condition is presented by the lower (two-loop in our case) contribution (one should remind that the one-loop effective potential is not zero but does not depends on $\mu$), we can obtain solutions for the complete effective potential up to the desired order in our arguments. So, effectively this methodology is a some kind of resummation of the whole effective potential where it is obtained as an expansion in $\log\frac{\phi^2}{\mu}$ and couplings rather than expansion in loops, i.e. our results allow to sum over loops in any given order of this new  expansion. In other words, within this methodology we present the alternative expansion (different from the loop one) of the complete effective action. 

Due to the nontrivial renormalization group functions, we can obtain the effective potential by imposing that $V_{eff}$ have to satisfy the RGE 
\begin{eqnarray}\label{rge1}
\left[\mu\frac{\partial}{\partial\mu}+\beta_\lambda\frac{\partial}{\partial\lambda}-\gamma_\phi \phi\frac{\partial}{\partial\phi}\right]V_{eff}(\phi;\mu,\lambda)=0~, 
\end{eqnarray}
where $\mu$ is a mass scale introduced by the regularization.

On general grounds, the effective superpotential $V_{eff}$ can be written as
\begin{eqnarray}\label{veffKF}
V_{eff}=\mathcal{K}(\phi)+\mathcal{F}(\phi,D^2\phi,\cdots)~,
\end{eqnarray}
where the superpotential $\mathcal{K}(\phi)$ is a function of $\phi$ but not of its derivatives $D^2\phi$, while $\mathcal{F}(\phi,D^2\phi,\cdots)$ is a function of $\phi$ and at least one derivative $D^2\phi$.

In order to use the RGE to compute the effective superpotential $V_{eff}$, we shall use the following ansatz for $\mathcal{K}(\phi)$ and $\mathcal{F}(\phi,D^2\phi)$:
\begin{eqnarray}
\mathcal{K}(\phi)&=&-\frac{1}{4}\phi^4~S_k(L)~,\label{ansatzk}\\
\mathcal{F}(\phi,D^2\phi)&=&-\frac{1}{2}\phi D^2\phi~S_f(L)~,\label{ansatzf}
\end{eqnarray}
where
\begin{eqnarray}\label{log1}
L=\log\left(\frac{\phi^2}{\mu}\right)~,
\end{eqnarray}
and 
\begin{eqnarray}\label{eq2S}
S(L)=A(\lambda)+B(\lambda)L+C(\lambda)L^2+D(\lambda)L^3+\cdots~.
\end{eqnarray}
The coefficients of $L$ ($A(\lambda)$, $B(\lambda)$, etc...) in the above equation are taken to be power series of the coupling constant $\lambda$.  

It is easy to see that some partial derivatives in the RGE (\ref{rge1}), by the use of (\ref{log1}), can be identified as derivatives of $L$, i.e., $\partial_L=-\mu\partial_\mu=\frac{1}{2}\phi\partial_\phi$. Thus, the RGE (\ref{rge1}) can be written in a convenient form as
\begin{eqnarray}\label{rge2}
&&\frac{1}{2}\phi D^2\phi\left[-\left(1-2\gamma_\phi\right)\frac{\partial}{\partial L}+\beta_\lambda\frac{\partial}{\partial\lambda}+\gamma_\phi\right]S_f(L)\nonumber\\
&&+
\frac{\phi^4}{4}\left[-\left(1-2\gamma_\phi\right)\frac{\partial}{\partial L}+\beta_\lambda\frac{\partial}{\partial\lambda}+4\gamma_\phi\right]S_k(L)=0~. 
\end{eqnarray}

Since each part of the above equation has to vanish independently, our task is to look for the functions $S_f(L)$ and $S_k(L)$ that satisfy 
\begin{eqnarray}\label{rge3}
&&\left[-\left(1-2\gamma_\phi\right)\frac{\partial}{\partial L}+\beta_\lambda\frac{\partial}{\partial\lambda}+\gamma_\phi\right]S_f(L)=0~,\label{eqsf}\\
&&\left[-\left(1-2\gamma_\phi\right)\frac{\partial}{\partial L}+\beta_\lambda\frac{\partial}{\partial\lambda}+4\gamma_\phi\right]S_k(L)=0~. \label{eqsk}
\end{eqnarray}

In general, the equations for $S(L)$ can be solved order by order in a series expansion of $L$ and the coupling constants. To compute the superpotentials $\mathcal{F}_{eff}$ and $\mathcal{K}_{eff}$, we insert the ansatz (\ref{eq2S}) for $S(L)$ into (\ref{rge3}), obtaining 
\begin{eqnarray}
&&-\left(1-2\gamma_{_\Phi}\right)\left[B(\lambda)+2C(\lambda)L+\cdots\right]+\beta_\lambda\frac{\partial}{\partial\lambda}\left[A(\lambda)+B(\lambda)L+\cdots\right]\nonumber\\
&&+\alpha_{_{F,K}}\gamma_{_\Phi}\left[A(\lambda)+B(\lambda)L+\cdots\right]=0~, \label{rge4}
\end{eqnarray}
where $\alpha_{_{F,K}}$ is the constant factor assuming the values $1$ or $4$, according to Eqs. (\ref{eqsf}) or (\ref{eqsk}), respectively.

The resulting expressions are organized by orders of $L$, obtaining a series of equations, that can be solved order by order in the coupling constant, i.e., the functions $A(\lambda)$, $B(\lambda)$, etc, are expanded as a power series of $\lambda$, 
\begin{eqnarray}
A(\lambda)&=&A^{(0)}+A^{(1)}\lambda+A^{(2)}\lambda^2+\cdots\nonumber\\
B(\lambda)&=&B^{(0)}+B^{(1)}\lambda+B^{(2)}\lambda^2+\cdots
\end{eqnarray}
and so on.

In the next section, we solve the equations for $S(L)$ for the simple case ${\cal N}=1$ Wess-Zumino model, using the renormalization group functions found in~\cite{Maluf:2012ie}. 

\section{${\cal{N}}=1$ Wess-Zumino model in the three-dimensional spacetime}\label{sec_wz}

The ${\cal{N}}=1$ Wess-Zumino model in the three-dimensional spacetime, with signature $(-,+,+)$, is defined by the following action
\begin{eqnarray}\label{wzeq01}
S&=& \int{d^3x~d^2\theta~}\Big{\{}\frac{1}{2}\Phi D^2\Phi +\frac{\lambda}{4!}\Phi^4\Big{\}}.
\end{eqnarray}

The real superfield $\Phi$ is expanded in a Taylor series in the Grassmannian variable as
\begin{eqnarray}\label{wzcomponents}
\Phi(x,\theta)=\varphi(x)+\theta^{\alpha}\psi_{\alpha}(x)-\theta^2F(x),
\end{eqnarray}
\noindent where $\varphi$ and $F$ are real scalar fields and $\psi$ is a two component Majorana fermion. Integrating (\ref{wzeq01}) over the Grassmannian variables, we obtain the action in terms of the component fields:
\begin{eqnarray}\label{wzeq01a}
S&=& \int{d^3x}\Big{\{}\frac{1}{2}\varphi \Box\varphi +\frac{i}{2}\psi^\alpha{\partial_{\alpha}}^\beta \psi_\beta+\frac{F^2}{2}
+\frac{\lambda}{2}\varphi^2\psi^2+\frac{\lambda}{3!}F\varphi^3\Big{\}}.
\end{eqnarray}

The auxiliary field $F$ can be eliminated from the action by the use of its equation of motion, $F=-\frac{\lambda}{3!}\varphi^3$, revealing the physical content of the model
\begin{eqnarray}\label{wzeq01b}
S&=& \int{d^3x}\Big{\{}\frac{1}{2}\varphi \Box\varphi +\frac{i}{2}\psi^\alpha{\partial_{\alpha}}^\beta \psi_\beta
+\frac{\lambda}{2}\varphi^2\psi^2
+\frac{\lambda^2}{72}\varphi^6\Big{\}}.
\end{eqnarray}

We will make use of the RGE to evaluate the effective superpotential. To do this, let us dislocate the superfield $\Phi(z)\rightarrow\Phi(z)+\phi(\theta)$, where $\phi(\theta)=(\varphi_{cl}-\theta^2 F_{cl})$ is a constant background superfield. In terms of the background superfield, the action (\ref{wzeq01}) can be cast as  
\begin{eqnarray}\label{wzeq2}
S&=&\int{d^3x~d^2\theta~}\Big{\{}\frac{1}{2}\Phi \left(D^2+\frac{\lambda\phi^2}{2}\right)\Phi 
+\frac{\lambda}{4!}\Phi^4+\nonumber\\ &+&\frac{\lambda}{3!}\phi\Phi^3
+\frac{1}{2}\left(D^2\phi+\frac{\lambda\phi^3}{3}\right)\Phi
+\frac{1}{2}\phi D^2\phi+\frac{\lambda}{4!}\phi^4\Big{\}}.
\end{eqnarray}



The Wess-Zumino model possesses nontrivial beta function $\beta(\lambda)$ and anomalous dimension $\gamma_{\Phi}$ renormalization group functions, see f.e. in \cite{Maluf:2012ie}, therefore the effective superpotential can be evaluated once $V_{eff}$ have to satisfy the RGE (\ref{rge1}). Since the theory requires renormalization, due to logarithmic UV divergences, the effective superpotential will exhibit a dependence on the logarithm of the classical superfield $\phi$, i.e., the effective superpotential should be dependent on $L$ (\ref{log1}), which we will use to construct our ansatz.

Let us start to compute the superpotential $\mathcal{K}_{eff}$ by solving the Eq.(\ref{eqsk}). We can use power series in $\lambda$ as the ansatz for the coefficients of $L$ in (\ref{eq2S}),
\begin{eqnarray}\label{coef1}
A_k=\sum_{n=1}^\infty a_n\lambda^n,\nonumber\\
B_k=\sum_{n=1}^\infty b_n\lambda^n, \nonumber\\
C_k=\sum_{n=1}^\infty c_n\lambda^n,
\end{eqnarray}
and so on. 

It is important to note that the leading logs contribution to $\mathcal{K}_{eff}$ can be written as
\begin{equation}\label{wzkeff}
\mathcal{K}_{eff}=\frac{1}{4!}\phi^4 \left[ \sum_{n=0}^\infty C_n^{ll}\lambda^{2n+1}L^n + \sum_{n=1}^\infty C_n^{Nll}\lambda^{2n+3}L^n +\cdots \right],
\end{equation}
so, taking into account only the leading logs, presented by the first term of this expression, with using the power series (\ref{coef1}) in (\ref{eq2S}) and substituting (\ref{eq2S}) into (\ref{eqsk}), we find the relations between the coefficients of (\ref{coef1}) looking as follows
\begin{eqnarray}\label{wzrelations}
 b_3&=&\frac{(4\gamma_\Phi\lambda+\beta_\lambda)}{\lambda^3}a_1,\nonumber\\
 c_5&=&\frac{(4\gamma_\Phi\lambda+2\beta_\lambda)}{\lambda^3}b_3=\frac{(4\gamma_\Phi\lambda+2\beta_\lambda)(4\gamma_\Phi\lambda+\beta_\lambda)}{2\lambda^6}a_1,\nonumber\\
 d_7&=&\frac{(4\gamma_\Phi\lambda+3\beta_\lambda)}{3\lambda^3}c_5=\frac{(4\gamma_\Phi\lambda+3\beta_\lambda)(4\gamma_\Phi\lambda+2\beta_\lambda)(4\gamma_\Phi+\beta_\lambda)}{3!\lambda^9}a_1,
\end{eqnarray}
and so on.

From a direct inspection of the above relations, we can find that the coefficients of the leading logs series obey the following recurrence relation,
\begin{equation}\label{wzrec_rel}
 C_n^{ll}=\frac{((2n-1)\beta_\lambda+4\gamma_\Phi\lambda}{n\lambda^3}C_{n-1}^{ll}.
\end{equation}
Here we fix the coefficient $C_0^{ll}=a_1=1$ to obtain the classical superpotential ${\mathcal K}(\phi)=\frac{\lambda}{4!}\phi^4$ at lowest order in $\lambda$.

Using the renormalization group functions given in the literature~\cite{Maluf:2012ie}, $\beta_\lambda=\frac{5\lambda^3}{24\pi^2}$ and $\gamma_\Phi=\frac{\lambda^2}{192\pi^2}$,  substituting (\ref{wzrec_rel}) into (\ref{wzkeff}) and performing the sum, we obtain
\begin{equation}\label{wzkeff_sum}
 \mathcal{K}_{eff}=\frac{\lambda}{4!}\phi^4 \left(\frac{\phi^2}{\mu}\right)^{\frac{11 \lambda^2}{48 \pi^2}}~.
 \end{equation}

Following the same steps, the effective superpotential $\mathcal{F}_{eff}$ can be evaluated as
\begin{equation}\label{wzfeff_sum}
 \mathcal{F}_{eff}=\frac{1}{2}\phi D^2\phi \left(\frac{\phi^2}{\mu }\right)^{\frac{41 \lambda^2}{192 \pi^2}}~.
 \end{equation}

Finally, the complete leading log effective superpotential can be cast as
\begin{equation}\label{wzveff_sum}
 V_{eff}=-\frac{1}{2}\left(\phi D^2\phi+\frac{\lambda}{12} \phi^4 \left(\frac{\phi^2}{\mu }\right)^{\frac{\lambda^2}{64 \pi^2}}\right) \left(\frac{\phi^2}{\mu }\right)^{\frac{41 \lambda^2}{192 \pi^2}}~.
 \end{equation}
 
In order to study the properties of the vacuum, let us write the effective potential $U_{eff}$ in terms of the component fields. It is obtained through integrating of $V_{eff}$ over the Grassmannian variable like in (\ref{veff_comp1}). Thus, we find
\begin{eqnarray}\label{+2}
U_{eff}=\int{d^2\theta}~V_{eff}=-\frac{F_{cl}^2}{2} \left(\frac{\varphi_{cl}^2}{\mu}\right)^{\frac{41 \lambda^2}{192\pi^2}}\left(1+\frac{41\lambda^2}{96\pi^2} \right)
-\frac{\lambda}{3!}F_{cl}\varphi_{cl}^3\left(\frac{\varphi_{cl}^2}{\mu}\right)^{\frac{11 \lambda^2}{48 \pi^2}}\left(1+\frac{11\lambda^2}{144\pi^2}\right).
\end{eqnarray}

The auxiliary field $F_{cl}$ can be eliminated by the use of its equation of motion, 
\begin{eqnarray}
F_{cl}=-\frac{\lambda  \left(144 \pi ^2+\lambda ^2\right) \varphi_{cl}^3}{9 \left(96 \pi ^2+41 \lambda^2\right)},
\end{eqnarray}
resulting in the physical effective potential
\begin{eqnarray}\label{+3}
U_{eff}=\frac{\left(144 \pi ^2 \lambda +\lambda ^3\right)^2}{15552 \pi^2 \left(96 \pi^2+41 \lambda^2\right)}
\varphi_{cl}^6 \left(\frac{\varphi_{cl}^2}{\mu}\right)^{\frac{11 \lambda^2}{48 \pi^2}}~.
\end{eqnarray}
This effective potential is well defined for all $\varphi_{cl}$ if $\lambda^2>-\frac{144\pi^2}{11}$, which is especially interesting since it is well defined for a real and perturbative coupling constant $\lambda$.

The conditions minimizing the effective potential are
\begin{eqnarray}\label{eq10}
&&\frac{dU_{eff}}{d\varphi_{cl}}\Big{|}_{\varphi_{cl}=\varphi_0}=
\frac{\left(144 \pi^2+11 \lambda^2\right) \left(144 \pi^2 \lambda +\lambda^3\right)^2}{373248 \pi^4 \left(96 \pi^2+41 \lambda^2\right)}\varphi_0^5 \left(\frac{\varphi_0^2}{\mu }\right)^{\frac{11 \lambda^2}{48 \pi^2}}=0~,\\
&&\frac{d^2U_{eff}}{d\varphi_{cl}^2}\Big{|}_{\varphi_{cl}=\varphi_0}
=\frac{\left(120 \pi^2+11 \lambda^2\right) \left(144 \pi^2+11 \lambda^2\right) \left(144 \pi^2 \lambda +\lambda^3\right)^2}{8957952 \pi ^6 \left(96 \pi ^2+41 \lambda ^2\right)}\varphi_0^4 \left(\frac{\varphi_0^2}{\mu}\right)^{\frac{11 \lambda^2}{48 \pi^2}}
>0~.
\end{eqnarray}
Keeping in mind the condition $\lambda^2>-144\pi^2/11$, we see that $\varphi_{cl}=\varphi_0=0$ is the minimum of the effective potential. Although the second derivative $\frac{d^2U_{eff}}{d\varphi_{cl}^2}$ also vanishes at $\varphi_0=0$, we see that the function $\frac{dU_{eff}}{d\varphi_{cl}}$ changes its sign from negative to positive one exactly at $\varphi_{cl}=0$, characterizing this point as a local minimum.

Since the minimum of the effective potential is exactly $\varphi_{cl}=0$, there is no generation of mass in the model. Just as occurs the purely scalar model in four dimensional spacetime~\cite{Coleman:1973jx}, the dynamical generation of a mass scale $\mu$ due to radiative corrections does not generate mass for the bosonic superfield in the present model.  

In the next sections, we deal with an extended supersymmetric gauge theory, the ${\cal N}=2$ Chern-Simons-matter model, using the renormalization group functions found earlier in~\cite{Avdeev:1991za}.

\section{The ${\cal N}=2$ Chern-Simons-matter model in ${\cal N}=1$ superspace}\label{n2csm}

The classical action for the ${\cal N}=2$ Chern-Simons-matter model in ${\cal N}=1$ superspace is given by
\begin{eqnarray}\label{ceq1}
S&=&\int{d^3x~d^2\theta~}\Big{\{}\frac{1}{2}\Gamma^{\alpha}W_{\alpha}
-\frac{1}{2}\overline{\nabla^{\alpha}\Phi}\nabla_{\alpha}\Phi
+\frac{g^2}{4}(\bar\Phi\Phi)^2\Big{\}},
\end{eqnarray}

\noindent
where  $W^{\alpha}=(1/2)D^{\beta}D^{\alpha}\Gamma_{\beta}$ is the gauge superfield strength and $\nabla^{\alpha}=(D^{\alpha}-ig\Gamma^{\alpha})$ it is the gauge supercovariant derivative. This model exhibit superconformal and gauge invariances at classical level.  

Our aim in this work is to use the RGE to compute the effective superpotential as discussed in the previous section. To do this, let us shift the superfields $\bar\Phi$ and $\Phi$ by the classical real background superfield $\phi(\theta)$ as 
\begin{eqnarray}\label{ceq5}
\bar\Phi &=&\frac{1}{\sqrt{2}}\left(\Phi_1+\phi-i\Phi_2\right)\nonumber\\
\Phi &=&\frac{1}{\sqrt{2}}\left(\Phi_1+\phi+i\Phi_2\right)~,
\end{eqnarray}
where $\Phi_1$ and $\Phi_2$ are real quantum matter superfields.

In terms of the background superfield, the action (\ref{ceq1}) can be written as
\begin{eqnarray}\label{ceq6}
S&=&\int{d^3x~d^2\theta~}\Big{\{}\frac{1}{2}\Gamma^{\alpha}W_{\alpha}-\frac{g^2\phi^2}{4}\Gamma^{\alpha}\Gamma_{\alpha}
-\frac{g\phi}{2}D^{\alpha}\Gamma_{\alpha}\Phi_2
+\frac{1}{2}\Phi_1\left(D^2+\frac{3g^2\phi}{4}\right)\Phi_1
+\frac{1}{2}\Phi_2\left(D^2+\frac{g^2}{2}\right)\Phi_2 \nonumber\\ 
&+&\frac{1}{2}\phi D^2\phi +\frac{g^2}{16}\phi^4
+\frac{g}{2}D^{\alpha}\Phi_2 \Gamma_{\alpha}\Phi_1
-\frac{g}{2}D^{\alpha}\Phi_1 \Gamma_{\alpha}\Phi_2
-\frac{g^2}{2}(\Phi_1^2+\Phi_2^2)\Gamma^2
-g^2\phi \Phi_1 \Gamma^2\nonumber\\
&+&\frac{g^2}{16}(\Phi_1^2+\Phi_2^2)^2
+\frac{g^2}{4} \phi \Phi_1(\Phi_1^2+\Phi_2^2)-dD^{\alpha}\phi  \Phi_2 \Gamma_{\alpha}
+\left(\frac{g^2}{4} \phi^3+D^2\phi\right)\Phi_1 \Big{\}}~.
\end{eqnarray}

The mixing between $\Gamma^{\alpha}$ and $\Phi_2$ superfields is eliminated by using an $R_{\xi}$ gauge fixing, $\mathcal{F}_{G}=(D^{\alpha}\Gamma_{\alpha}+\alpha g\phi \Phi_2/2)$. The corresponding gauge fixing and Faddeev-Popov actions are given by
\begin{eqnarray}\label{ceq6a}
S_{GF+FP}&=&\int{d^3x~d^2\theta~}\Big[\frac{1}{2\alpha}
(D^{\alpha}\Gamma_{\alpha}+\alpha\frac{g\phi }{2}\Phi_2)^2
+\bar{C}D^2C+\frac{\alpha}{4}g^2\phi^2\bar{C}C+\frac{\alpha}{4}{g^2\phi }\bar{C}\Sigma C\Big]~.
\end{eqnarray}

Notice that the mass terms of the gauge $\Gamma^{\alpha}$ and matter $\Phi_1$ superfields are identified as $M_{\Gamma}=\frac{g^2\phi^2}{4}$ and $M_{\Phi_1}=\frac{3g^2\phi^2}{4}$, respectively.

\section{Renormalization group improvement of the effective superpotential}\label{rge_veff}

In this section we apply the method described in the section \ref{RGE-superspace} to compute the effective superpotential of the ${\cal N}=2$ Chern-Simons-matter model. We use the renormalization group functions found earlier in~\cite{Avdeev:1991za}, where the only non-trivial function is the anomalous dimension of the matter superfield $\gamma_{_\Phi}=-\frac{(N+2)g^4}{4}=-\frac{(N+2)\eta^2}{4}$, with $N$ being the number of flavors of the matter superfields. For convenience, we will use $\eta=g^2$ to count the orders of coupling constant.

Considering the effective potential given in the Eq.(\ref{veffKF}), and the ansatz Eqs. (\ref{ansatzk}) and (\ref{ansatzf}), the RGE for $V_{eff}$, Eqs. (\ref{eqsf}) and (\ref{eqsf}), are given by
\begin{eqnarray}
&&\left[-\left(1-2\gamma_{_\Phi}\right)\frac{\partial}{\partial L}+\gamma_{_\Phi}\right]S_f(L)=0~,\label{eqsf1}\\
&&\left[-\left(1-2\gamma_{_\Phi}\right)\frac{\partial}{\partial L}+4\gamma_{_\Phi}\right]S_k(L)=0~, \label{eqsk1}
\end{eqnarray}
where we have used that $\beta(g)=\beta(\eta)=0$.

Now, let us compute the $\mathcal{K}_{eff}$ superpotential first. Inserting the ansatz (\ref{eq2S}) for $S(L)$ into (\ref{eqsk1}), we have 
\begin{eqnarray}
&&-\left(1-2\gamma_{_\Phi}\right)\left[B_k(\eta)+2C_k(\eta)L+\cdots\right]+4\gamma_{_\Phi}\left[A_k(\eta)+B_k(\eta)L+\cdots\right]=0~. \label{eqsk2}
\end{eqnarray}

We can organize the resulting expression by orders of $L$, obtaining a series of equations, of which we write down  the first three:
\begin{eqnarray}
&&-\left(1-2\gamma_{_\Phi}\right)B_k(\eta)+4\gamma_{_\Phi}A_k(\eta)=0~;\label{eq:orderL0-k}
\end{eqnarray}
\begin{eqnarray}
&&-2\left(1-2\gamma_{_\Phi}\right)C_k(\eta)+4\gamma_{_\Phi}B_k(\eta)=0\thinspace;\label{eq:orderL1-k}
\end{eqnarray}
and
\begin{eqnarray}
&&-3\left(1-2\gamma_{_\Phi}\right)D_k(\eta)+4\gamma_{_\Phi}C_k(\eta)=0\thinspace.\label{eq:orderL2-k}
\end{eqnarray}

The functions $A(\eta)$, $B(\eta)$, etc..., are expanded as a series in power of $\eta$, 
\begin{eqnarray}
A(\eta)&=&A^{(0)}+A^{(1)}\eta+A^{(2)}\eta^2+\cdots\nonumber\\
B(\eta)&=&B^{(0)}+B^{(1)}\eta+B^{(2)}\eta^2+\cdots\nonumber
\end{eqnarray}
and so on.

Organizing the above equations order by order in powers of $\eta$, we find
\begin{eqnarray}
B^{\left(3\right)}&=&4\gamma_{_\Phi}^{\left(2\right)} A^{\left(1\right)};\label{p1}\\
C^{\left(5\right)}&=&\frac{1}{2}4\gamma_{_\Phi}^{\left(2\right)} B^{\left(3\right)}=\frac{1}{2}\left(4\gamma_{_\Phi}^{\left(2\right)}\right)^2A^{\left(1\right)}\label{p2}\\
D^{\left(7\right)}&=&\frac{1}{3}4\gamma_{_\Phi}^{\left(2\right)} C^{\left(5\right)}=\frac{1}{3!}\left(4\gamma_{_\Phi}^{\left(2\right)}\right)^3A^{\left(1\right)}\label{p3}. 
\end{eqnarray}

Writing the effective superpotential $\mathcal{K}_{eff}$ as
\begin{eqnarray}\label{veffsum}
\mathcal{K}_{eff}=-\frac{\phi^4}{16} \sum_{n=0}^{\infty} C_n^{ll}\eta^{2n+1}L^n~,
\end{eqnarray}
we find from Eqs.(\ref{p1},\ref{p2},\ref{p3}) the following recurrence relation
\begin{eqnarray}\label{cll}
C_n^{ll}=\left(\frac{4\gamma_{_\Phi}^{\left(2\right)}}{n~\eta^2}\right) C_{n-1}^{ll},
\end{eqnarray}
\noindent where we identify $C_0^{ll}=A^{(1)}=1$, $C_1^{ll}=B^{(3)}$, $C_2^{ll}=C^{(5)}$, $C_3^{ll}=D^{(7)}$ and so on.

Inserting (\ref{cll}) into  (\ref{veffsum}) and performing the sum, we obtain the following effective superpotential  
\begin{eqnarray}\label{keffsum01}
\mathcal{K}_{eff}= -\frac{\eta}{16}\phi^4 \left(\frac{\phi^2}{\mu}\right)^{-(N+2)\eta^2}=-\frac{g^2}{16}\phi^4\left(\frac{\phi^2}{\mu}\right)^{-(N+2)g^4}.
\end{eqnarray}


Following the same steps above, the $\mathcal{F}_{eff}$ can be cast as
\begin{eqnarray}\label{feffsum01}
\mathcal{F}_{eff}= -\frac{1}{2}\phi D^2\phi \left(\frac{\phi^2}{\mu}\right)^{-(N+2)\eta^2/4}
=-\frac{1}{2}\phi D^2\phi \left(\frac{\phi^2}{\mu}\right)^{-(N+2)g^4/4}.
\end{eqnarray}


Therefore, the full effective superpotential is given by
\begin{eqnarray}\label{veffsum01}
V_{eff}= -\frac{g^2}{16}\phi^4\left(\frac{\phi^2}{\mu}\right)^{-(N+2)g^4}-\frac{1}{2}\phi D^2\phi \left(\frac{\phi^2}{\mu}\right)^{-(N+2)g^4/4}.
\end{eqnarray}

Considering $\phi(\theta)=(\varphi_{cl}-\theta^2F_{cl})$ as the $\theta$ expansion of the background superfield, the effective potential $U_{eff}$, is given by
\begin{eqnarray}\label{veffsum02}
U_{eff}&=&\int{d^2\theta}~V_{eff}\nonumber\\
&=& -\left[1-\frac{(N+2)g^4}{2}\right]\left(\frac{\varphi_{cl}^2}{\mu}\right)^{-(N+2)g^4/4}
\left[\frac{F_{cl}^2}{2}+\frac{g^2}{4}F_{cl}\varphi_{cl}^3 \left(\frac{\varphi_{cl}^2}{\mu}\right)^{-3(N+2)g^4/4}\right]\nonumber\\
&=&\frac{\eta^2}{64} \left[2-(N+2)\eta^2\right] \varphi_{cl}^6 \left(\frac{\varphi_{cl}^2}{\mu }\right)^{-\frac{7}{4} (N+2)\eta^2}~,
\end{eqnarray}
where we have used the equation of motion of the auxiliary field, $F_{cl}=-\frac{1}{4} \eta \varphi_{cl}^3 \left(\frac{\varphi_{cl}^2}{\mu }\right)^{-\frac{3}{4} (N+2) \eta^2}$, to eliminate it from the effective potential.

In order to study a possible spontaneous generation of mass induced by radiative corrections, let us evaluate the minimum of the effective potential  $U_{eff}$. The conditions that minimize the $U_{eff}$ are given by:
\begin{eqnarray}\label{eq3dd}
&&\frac{d U_{eff}}{d\varphi_{cl}} = \frac{\eta^2}{128} \left(-2+(N+2)\eta^2\right) \left(-12+7 (N+2)\eta^2\right) \varphi_{cl}^5 \left(\frac{\varphi_{cl}^2}{\mu }\right)^{-\frac{7}{4} (N+2)\eta^2}=0~,\label{gap01}
\\
&& \frac{d^2U_{eff}}{d \varphi_{cl}^2}\Big{|}_{\varphi_{cl}=\varphi_0}>0~.\label{massphi}
\end{eqnarray}

It is easy to see that the only possible solution to (\ref{gap01}) is to take the limit $\varphi_{cl}=0$, if and only if $\eta^2<\frac{10}{7(N+2)}$, otherwise the effective potential presents a vertical asymptote at $\varphi_{cl}=0$. Such solution preserves the ${\cal N}=2$ supersymmetry and no mass is dynamically generated by radiative corrections. 

The second derivative, Eq.(\ref{massphi}), is also vanishing in the minimum, but we can see that the function $dU_{eff}/d\varphi_{cl}$ changes its sign in the critical point, i.e., $dU_{eff}/d\varphi_{cl}<0$ for 
$\varphi_{cl}<0$ and $dU_{eff}/d\varphi>0$ for 
$\varphi_{cl}>0$, characterizing the critical point $\varphi=0$ as a local minimum, since $\eta^2<\frac{10}{7(N+2)}$.

Even though in the three-dimensional spacetime the classical minimum ($F_{cl}=0$) no longer needs be a minimum of the quantum effective potential, in the present model the supersymmetry is preserved by quantum corrections, at least to the leading logs. This result is similar to the scenario taking place in the three-dimensional Wess-Zumino model discussed in \cite{Maluf:2012ie}, where the authors computed the superpotential up to two-loop order.

\section{Summary}\label{summary}

In this work we explicitly demonstrated how to apply the Renormalization Group Equation for evaluating the effective superpotential in terms of a ${\cal N}=1$ background superfield in a three-dimensional spacetime. This method represents itself as a powerful technique allowing us to compute higher-order corrections on the base the two-loop Renormalization Group Functions. As an important example, due to non-trivial anomalous dimension of the matter superfield, we computed the complete leading log effective superpotential in the ${\cal N}=1$ Wess-Zumino model and the ${\cal N}=2$ supersymmetric Chern-Simons-matter theory and demonstrated that they have an expected form $\Phi^4\left[\frac{\Phi^2}{\mu}\right]^{\alpha}$ and $\Phi D^2\Phi\left[\frac{\Phi^2}{\mu}\right]^{\alpha}$. We discussed the vacuum properties of the quantum effective superpotential in such model, and showed that no mass is generated by radiative corrections, differently from what happens for the ${\cal N}=1$ version of the model~\cite{Ferrari:2010ex,Lehum:2010tt,Quinto:2014zaa,Gallegos:2011ux,Queiruga:2015xaa}. We argued that the supersymmetry is preserved by quantum corrections.

We expect that this method can be applied to other superfield models involving the Chern-Simons theory as an ingredient. These models will be studied in our next papers.

\vspace{.5cm}
{\bf Acknowledgements.} 
This work was partially supported by Conselho Nacional de Desenvolvimento Cient\'{\i}fico e Tecnol\'{o}gico (CNPq). A.C.L. has been partially supported by the CNPq project 307723/2016-0 and 402096/2016-9. The work by A.Yu.P. has been supported by CNPq project 303783/2015-0.

\end{document}